\documentclass[aps,prb,twocolumn,superscriptaddress,showpacs]{revtex4}
\usepackage{graphicx}
\usepackage{latexsym}
\usepackage{amssymb}
\usepackage{amsmath}
\usepackage{amsfonts}
\usepackage{bm}
\usepackage{multirow}
\usepackage{color}
\usepackage{comment}

\newcommand{\Tr}{\mathop{\mathrm{Tr}}}

\newcommand{\SU}{\mathrm{SU}}
\newcommand{\U}{\mathrm{U}}

\newcommand{\beq}{\begin{equation}}
\newcommand{\eeq}{\end{equation}}
\newcommand{\beqn}{\begin{eqnarray}}
\newcommand{\eeqn}{\end{eqnarray}}

\DeclareMathAlphabet{\mathbbold}{U}{bbold}{m}{n}

\def\Tr{\text{Tr}}

\def\SU{{\rm SU}}
\def\PSU{{\rm PSU}}
\def\U{{\rm U}}

\makeatletter
\newcommand\xleftrightarrow[2][]{%
\ext@arrow 9999{\longleftrightarrowfill@}{#1}{#2}}
\newcommand\longleftrightarrowfill@{%
\arrowfill@\leftarrow\relbar\rightarrow} \makeatother

\begin{document}

\title{Lieb-Schultz-Mattis Theorem and its generalizations from the Perspective of the \\ Symmetry Protected Topological phase}

\author{Chao-Ming Jian}
\affiliation{Kavli Institute of Theoretical Physics, Santa
Barbara, CA 93106, USA} \affiliation{ Station Q, Microsoft
Research, Santa Barbara, California 93106-6105, USA}

\author{Zhen Bi}
\affiliation{Department of Physics, University of California,
Santa Barbara, CA 93106, USA}

\author{Cenke Xu}
\affiliation{Department of Physics, University of California,
Santa Barbara, CA 93106, USA}

\begin{abstract}

We ask whether a local Hamiltonian with a featureless (fully
gapped and nondegenerate) ground state could exist in certain
quantum spin systems. We address this question by mapping the
vicinity of certain quantum critical point (or gapless phase) of
the $d-$dimensional spin system under study to the boundary of a
$(d+1)-$dimensional bulk state, and the lattice symmetry of the
spin system acts as an on-site symmetry in the field theory that
describes both the selected critical point of the spin system, and
the corresponding boundary state of the $(d+1)-$dimensional bulk.
If the symmetry action of the field theory is nonanomalous, $i.e.$
the corresponding bulk state is a trivial state instead of a
bosonic symmetry protected topological (SPT) state, then a
featureless ground state of the spin system is allowed; if the
corresponding bulk state is indeed a nontrivial SPT state, then it
likely excludes the existence of a featureless ground state of the
spin system. From this perspective we identify the spin systems
with SU($N$) and SO($N$) symmetries on one, two and three
dimensional lattices that permit a featureless ground state. We
also verify our conclusions by other methods, including an
explicit construction of these featureless spin states.

\end{abstract}


\maketitle

\section{Introduction}

The Lieb-Schultz-Mattis (LSM) theorem~\cite{LSM}, and its higher
dimensional generalizations~\cite{oshikawa,hastings} state that if
a quantum spin system defined on a lattice has odd number of
spin-1/2s per unit cell, then any local spin Hamiltonian which
preserves the spin and translation symmetry, cannot have a
featureless (gapped and nondegenerate) ground state. This implies
that any symmetry allowed Hamiltonian on the spin Hilbert space
defined above can only have the following possible scenarios: {\it
1.} its ground state spontaneously breaks either the spin symmetry
or the lattice symmetry, hence leads to degenerate ground states
and possible gapless Goldstone modes; {\it 2.} it has gapped and
degenerate ground states without breaking any symmetry, $i.e.$ its
ground state develops a topological order (the second possibility
can only happen in two and higher dimensional systems); {\it 3.}
its ground state has algebraic (power-law) correlation function of
physical quantities, and the spectrum is again gapless (this
scenario happens most often in $1d$ spin systems, while still
possible in higher dimensions).

On the other hand, there are lattice spin systems for which one
can very easily construct a local Hamiltonian with a featureless
ground state that preserves all the symmetry. One class of such
states are called the AKLT states~\cite{AKLT}, which can be
constructed for an integer spin chain in $1d$, the spin-2
antiferromagnet on the square lattice, and the spin-3/2
antiferromagnet on the honeycomb lattice, etc. Of course, these
systems violate the crucial ``odd number of spin-1/2s per unit
cell" assumption of the LSM theorem.

However, there are also some spin systems in the ``middle ground"
where the answers are not so clear. These systems do not meet the
key assumption of the LSM theorems, while a simple analogue of the
AKLT state mentioned above does not obviously exist. For example,
the honeycomb lattice has two sites per unit cell, thus a spin-1/2
system on the honeycomb lattice has even number of spin-1/2s per
unit cell, and hence there is no LSM theorem to exclude a
featureless ground state. But it has been a long standing problem
whether a featureless spin-1/2 state exists or not on the
honeycomb lattice. Another example is the spin-1 antiferromagnet
on the square lattice. Depending on the Hamiltonian, possible
states of this system include the N\'{e}el state which
spontaneously breaks the spin symmetry, and a nematic type of
valence bond solid state which breaks the lattice rotation
symmetry, etc. But the existence of a featureless state is not
obvious. However, recent progresses indicate that featureless
states do exist in these two ``middle ground" examples mentioned
above~\cite{Kimchi,jian1,jian2}, with a more sophisticated
construction compared with the AKLT state.

Another seemingly very different subject is the symmetry protected
topological (SPT) state~\cite{wenspt,wenspt2}, which is a
generalization of topological insulators. By definition, the
ground state of the $(d+1)-$dimensional bulk of a SPT phase must
be gapped and nondegenerate, while its $d-$dimensional boundary
state must be either gapless or degenerate, as long as certain
symmetries are preserved. In the last few years, the
classification of bosonic SPT states with on-site internal
symmetries has been well
understood~\cite{wenspt,wenspt2,senthilashvin,xuclass,kapustin1,kapustin4,freed1,freed2,wenso,bixu}.
%and the possible boundary states of these SPT states were also
%well-studied~\cite{senthilashvin,xuclass}.
The $d-$dimensional boundary of a $(d+1)-$dimensional SPT state,
just like those $d-$dimensional spin systems where the LSM theorem
applies, cannot be trivially gapped. The key difference between
these two systems is that, the former is (usually) protected by an
on-site symmetry, while the latter is protected by the spin and
lattice symmetries together. However, the fact that neither system
permits a featureless state suggests that we can potentially
formulate both systems in a similar way. The connection to $3d$
bulk SPT states has been exploited in order to understand the
fractional excitations of $2d$ topological orders with both spin
and lattice symmetries~\cite{menglsm}.

%The reason the boundary of a SPT state cannot be trivially gapped,
%is because it is ``anomalous", and cannot exist without the bulk.
%Thus to make connection between the LSM theorem and the SPT
%boundary state, roughly speaking we need to verify that the
%lattice symmetry of the spin system, which is crucial for the LSM
%theorem, once ``interpreted" as an on-site symmetry, becomes
%anomalous.

Since we are comparing two $d-$dimensional systems with very
different ultraviolet regularizations, their analogue can only be
made precise when both systems are tuned close to a point where a
low energy field theory description becomes available.
%Analysis in the vicinity of that
%point can already capture the anomaly, because anomaly is a
%universal feature which should exist in both ultraviolet and
%infrared descriptions.
Thus for our purpose, when we analyze a $d-$dimensional spin
system, we will first tune it to a critical point described by a
field theory, then interpret the lattice symmetry as an on-site
symmetry, and interpret the $d-$dimensional field theory as the
boundary state of a $(d+1)-$dimensional bulk. If the corresponding
$(d+1)-$dimensional bulk is a trivial state instead of a
nontrivial SPT state, then a featureless spin state must exist not
too far from that critical point in the phase diagram; if the
corresponding bulk is indeed a nontrivial SPT state, then it
highly suggests that a featureless spin ground state does not
exist.

However, the latter statement may not be necessarily true: if
around that selected critical point of the spin system the field
theory is formally equivalent to a SPT boundary state, it only
rules out the featureless spin state at the vicinity of that
critical point. But in principle a featureless state could be far
away from the critical point in the phase diagram, and hence
beyond the reach of the field theory.

In section II through V, we will discuss SU(N) and SO(N) systems
on a $1d$ chain, $2d$ square lattice, $2d$ honeycomb lattice, and
$3d$ cubic lattice respectively, by mapping them to the boundary
of $2d$, $3d$ and $4d$ bulk states. We will identify those spin
systems that permit a featureless ground state. For all of these
spin systems, we can explicitly construct a featureless tensor
product state that is an analogue of the AKLT state. Some examples
of these featureless states will be discussed in section VI. In
section VI we will also verify our conclusions by making
connection with a previous study on LSM theorem based on lattice
homotopy class~\cite{Po2017}.

\section{$1d$ spin chain}

\subsection{SU(2) spin-1/2 chain}

In this section we first discuss one dimensional spin chains with
SU(2) symmetry. The low energy physics of the Heisenberg
antiferromagnetic spin-1/2 chain with a SU(2) spin symmetry can be
captured by the following nonlinear sigma model in $(1+1)d$ with a
Wess-Zumino-Witten (WZW) term at level-1~\cite{affleck1986}: \beqn
\mathcal{S} = \int dx d\tau \ \frac{1}{g}(\partial_\mu \vec{n})^2
+ \frac{2\pi i}{\Omega_3} \int_0^1 du \ \epsilon_{abcd} n^a
\partial_\tau n^b
\partial_x n^c
\partial_u n^d, \label{o4wzw} \eeqn where $\vec{n}$ is a four component vector
with unit length, and $\Omega_3$ is the volume of $S^3$ with unit
radius. The physical meaning of $\vec{n}$ is that, $(n_1, n_2,
n_3)$ are the three component N\'{e}el order parameter, while $n_4
\sim \phi$ is the valence bond solid (VBS) order parameter. If
there is a SO(4) rotation symmetry of the four component vector
$\vec{n}$, the coupling constant $g$ will flow to a fixed point,
which corresponds to the SU(2)$_1$ conformal field
theory~\cite{Witten1984,KnizhnikZamolodchikov1984}. The SO(4)
symmetry becomes an emergent symmetry of the spin-1/2 Heisenberg
chain in the infrared: the N\'{e}el and VBS order parameter both
have the same scaling dimension $[\vec{n}] = 1/2$. The key
symmetry of the system, is the spin SU(2) symmetry, and the
translation symmetry. $(n_1, n_2, n_3)$ transforms as a vector
under spin SU(2), and $n_4 \sim \phi$ is a SU(2) singlet; and
under translation by one lattice constant, $T_x: \vec{n}
\rightarrow - \vec{n}$. The physical meaning of Eq.~\ref{o4wzw} is
the intertwinement between the N\'{e}el and VBS order parameter:
the domain wall of the VBS order parameter carries a spin-1/2.

%If one turns on SU(2) and translational invariant perturbations on
%the SU(2)$_1$ CFT, for instance $\lambda \mathbf{J}_L \cdot
%\mathbf{J}_R$, where $\mathbf{J}_L$ and $\mathbf{J}_R$ are the
%SU(2) currents of the left and right moving modes, then depending
%on the sign of $\lambda$, $\lambda$ is either marginally relevant
%or irrelevant. When $\lambda$ is marginally relevant, it leads to
%spontaneous translation symmetry breaking, and the VBS order
%parameter $n_4 \sim \phi$ develops a nonzero expectation value.

\begin{figure}[tbp]
\begin{center}
\includegraphics[width=220pt]{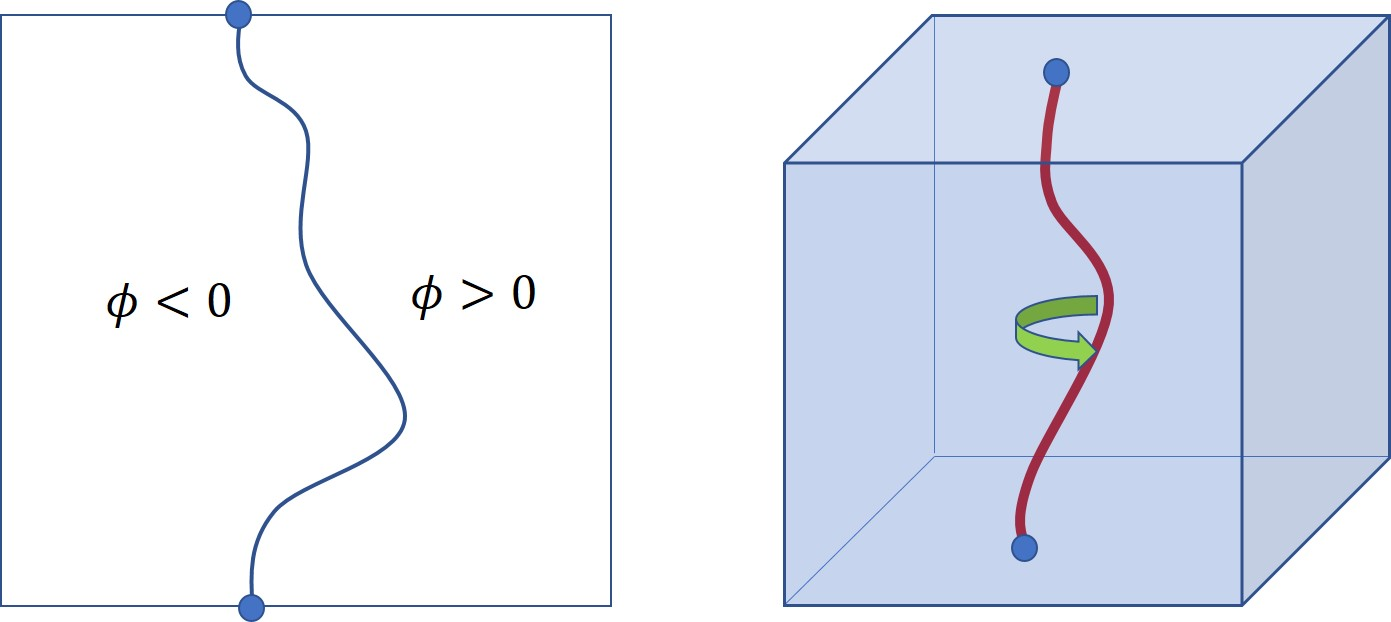}
\caption{$(a)$, the decorated domain wall construction of the $2d$
SPT state whose boundary is analogous to a SU($2N$) spin chain
with a LSM theorem. A $1d$ SPT state with PSU($2N$) symmetry is
decorated to each domain wall, and the domain wall terminates at
the boundary with a dangling projective representation of the
PSU($2N$) SPT state. $(b)$, the decorated vortex line construction
of the $3d$ SPT state whose boundary is analogous to a $2d$ spin
system either on the square or honeycomb lattice. Again, we
decorate each vortex line with a $1d$ SPT state. But when the $2d$
boundary is mapped to the square and honeycomb lattice spin
systems, the vortex line in the bulk has a $Z_4$ and $Z_3$
conservation, which must be compatible with the classification of
the $1d$ SPT state decorated on each vortex line in order to
guarantee a nontrivial $3d$ SPT. } \label{domain}
\end{center}
\end{figure}

The field theory Eq.~\ref{o4wzw} also describes the boundary of a
$2d$ bosonic SPT state with SO(3)$\times Z_2$
symmetry~\cite{xu2dspt,xusenthil}, where $Z_2$ acts as $\vec{n}
\rightarrow - \vec{n}$. This SPT state can be understood as the
decorated domain wall construction~\cite{chenluashvin}: we
decorate every $Z_2$ symmetry breaking domain wall in the $2d$
bulk with a Haldane phase with SO(3) symmetry
(Fig.~\ref{domain}$a$), then proliferate the $Z_2$ domain walls to
restore the $Z_2$ symmetry. The so-constructed phase in the bulk
is the desired SO(3)$\times Z_2$ SPT phase. And at the $1d$
boundary of the system, there is a spin-1/2 degree of freedom
localized at every $Z_2$ domain wall, which is also the boundary
state of the Haldane phase decorated at each $Z_2$ domain wall in
the bulk. This is consistent with the physics of the spin-1/2
chain.

This simple example demonstrates that the lattice translation
symmetry, once interpreted as an on-site symmetry in a field
theory, is equivalent to an ``anomalous" symmetry of the boundary
of a higher dimensional SPT state. And by definition the boundary
of a SPT state cannot be trivially gapped without degeneracy,
which is consistent with the LSM theorem of the spin-1/2
chain~\cite{LSM}.

Here we stress that, the $1d$ SPT phase decorated at a $Z_2$
domain wall must have a $Z_2$ classification as long as the
symmetry $G$ of the $1d$ SPT phase commutes with the $Z_2$, $i.e.$
two of the $1d$ SPT phases must fuse into a trivial state. One way
to see this is that, after gauging the $Z_2$ symmetry, the vison
($\pi-$flux introduced by the $Z_2$ gauge symmetry) preserves the
symmetry $G$ as long as $G$ commutes with $Z_2$, and the vison is
the boundary of the $1d$ decorated SPT state~\cite{chenluashvin}.
Since two visons fuse into a local excitation, the $1d$ SPT state
must have a $Z_2$ classification. But at a $Z_2^T$ (time-reversal)
domain wall one can decorate a lower-dimensional SPT phase with
(for example) $Z$ classification, because the anti-domain wall of
$Z_2^T$ is the time-reversal conjugate of a $Z_2^T$ domain wall,
which is automatically decorated with the ``inverse" state of the
SPT state decorated at the $Z_2^T$ domain wall~\footnote{The
authors thank Dung-Hai Lee for clarifying this important point for
us.}. This observation is consistent with many known facts about
SPT phases. For instance, in three dimensional space, there is a
$Z_2^T$ SPT which can be viewed as $Z_2^T$ domain walls decorated
with the $E_8$ invertible topological order~\cite{senthilashvin},
but there is no such decorated domain wall construction for $3d$
SPT phases with a $Z_2$ symmetry.

\subsection{spin chain with reduced symmetry}

Now one can exploit the connection between $1d$ spin chains and
the boundary of $2d$ SPT states even further, and consider a spin
chain with a reduced spin symmetry. For example we can start with
a spin-1/2 chain, and break the SO(3) spin symmetry down to its
subgroup $G \rtimes Z_2$, where $Z_2$ is the spin $\pi-$rotation
$S^z \rightarrow - S^z$, $S^y \rightarrow - S^y$, and $G$ is a
subgroup of the inplane U(1) spin symmetry. Whether the spin chain
can be featureless or not, is equivalent to the problem of whether
the corresponding bulk state with $(G \rtimes Z_2) \times Z_2$
symmetry is a nontrivial SPT state or not; and based on the
``decorated domain wall" picture mentioned above, this again is
equivalent to the problem of whether the $1d$ $Z_2$ domain wall is
a nontrivial $1d$ SPT state with $G \rtimes Z_2$ symmetry or not,
and if it is indeed a nontrivial SPT, whether it has a $Z_2$
classification.

Now we can look up the classification in
Ref.~\onlinecite{wenspt,wenspt2}. For example, when $G = Z_{2n+1}$
with integer $n$, since there is no nontrivial $1d$ SPT state with
$Z_{2n+1} \rtimes Z_2$ symmetry, the bulk SO(3)$\times Z_2$ SPT
state discussed previously must be trivialized by reducing the
SO(3) spin symmetry down to $Z_{2n+1} \rtimes Z_2$, thus its
boundary can in principle be gapped and nondegenerate. This
observation already gives us a meaningful conclusion:

{\it A spin chain with translation and $(Z_{2n+1} \rtimes Z_2)$
spin symmetry can have a featureless ground state.}

%In fact, the reason there is no $1d$ SPT state with $Z_{2n+1}
%\rtimes Z_2$ symmetry is simply that there is no projective
%representation of $Z_{2n+1} \rtimes Z_2$ group. Then this implies
%that for the $Z_{2n+1} \rtimes Z_2$ spin group, there is no
%essential difference between ``integer" spin chain and
%``half-integer" spin chain. Thus it is not surprising that a $1d$
%spin chain with $Z_{2n+1} \rtimes Z_2$ symmetry always permits a
%featureless gapped state.

By contrast, for $G = U(1)$ or $Z_{2n}$, a nontrivial $1d$ SPT
state with $G \rtimes Z_2$ does exist, and it does have a $Z_2$
classification. Hence the Haldane phase with SO(3) spin symmetry
remains a nontrivial SPT state under the symmetry reduction to $G
\rtimes Z_2$. Thus the $2d$ bulk SPT state with $(G \rtimes Z_2)
\times Z_2$ remains nontrivial, and hence the $1d$ boundary cannot
be trivially gapped. This observation leads to the following
conclusion:

{\it A $1d$ spin-1/2 chain cannot have a featureless ground state,
even if we break the SU(2) spin symmetry down to $(Z_{2n} \rtimes
Z_2)$ symmetry.}

\subsection{SU($2N$) spin chain}

\label{su2nchain}

\begin{figure}[tbp]
\begin{center}
\includegraphics[width=180pt]{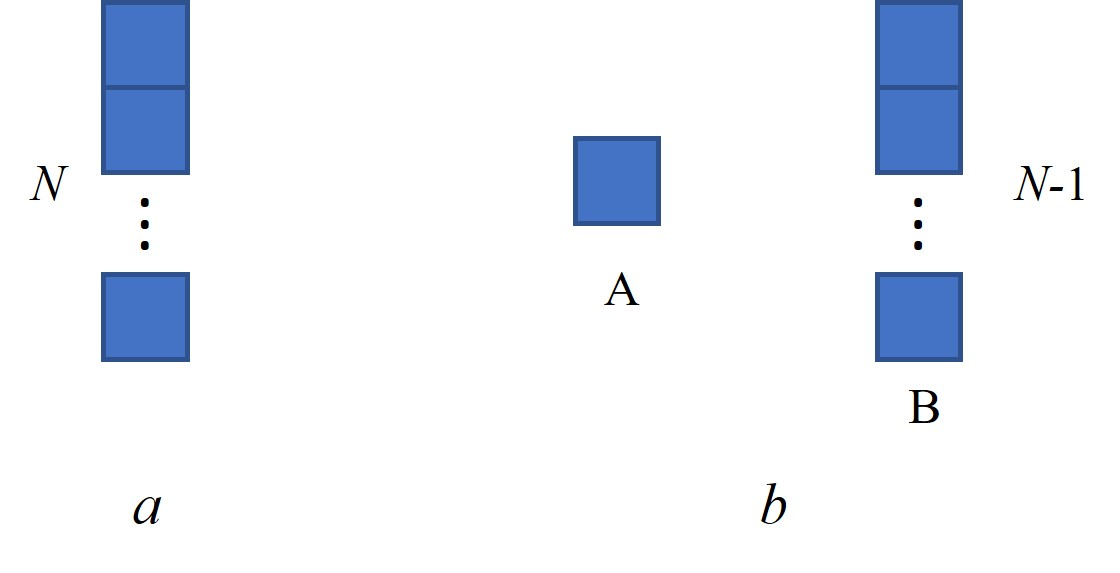}
\caption{$(a)$, the self-conjugate SU($2N$) spin representation on
each site considered in section~\ref{su2nchain}. $(b)$, For the
square, honeycomb and cubic lattice, we consider a SU($N$) spin
system with a fundamental representation (FR) on sublattice A, and
an anti-fundamental representation (AFR) on sublattice B. }
\label{rep}
\end{center}
\end{figure}

Now let's consider spin chains with higher spin symmetries. A
natural generalization of the spin-1/2 chain with translation
symmetry, is a SU($2N$) spin chain with self-conjugate
representation on each site (Young tableau with $N$ boxes in one
column, Fig.~\ref{rep}$a$). The analogue of the ``N\'{e}el" order
parameter of this SU($2N$) spin chain, is a $2N \times 2N$
Hermitian matrix order parameter $\mathcal{P}$, and it can be
represented in the form \beqn \mathcal{P} = V \Omega V^\dagger,
\;\;\Omega \equiv\! \left(
\begin{array}{cccc}
\mathbf{1}_{N \times N} & \mathbf{0}_{ N \times N} \\ \\
\mathbf{0}_{N \times N} & \;\;\mathbf{-1}_{N \times N}
\end{array}
\right) \;\label{P}\eeqn where $V$ is a SU($2N$) matrix. All the
configurations of $\mathcal{P}$ belong to the Grassmanian manifold
$\mathcal{M} = U(2N)/[U(N) \times
U(N)]$~\cite{affleck,sachdevread}. To see that $\mathcal{P}$ is a
natural generalization of the ordinary SU(2) N\'{e}el order
parameter, we can take $N=1$, then this Grassmanian is precisely
$S^2$, which is the manifold of the ordinary SU(2) N\'{e}el order
parameter. We can also define matrix order parameter $\mathcal{P}
= \vec{n}\cdot \vec{\sigma}$ for the SU(2) spin chain, where
$\vec{n}$ is the SU(2) N\'{e}el order parameter.

The effective field theory for the SU($2N$) spin chain described
above, can be written as~\cite{affleck}: \beqn \mathcal{S} = \int
dx d\tau \ \frac{1}{g} \mathrm{tr}[\partial_\mu \mathcal{P}
\partial_\mu \mathcal{P}] + \frac{\Theta}{16 \pi}
\epsilon_{\mu\nu} \mathrm{ tr}[ \mathcal{P}
\partial_\mu \mathcal{P} \partial_\nu \mathcal{P} ]. \label{PL}\eeqn This is
the analogue of the Nonlinear sigma model for the SU(2) spin
chain~\cite{Haldane1,Haldane2}, with a $\Theta-$term which comes
from the fact that for all $N$, the Grassmanian $\mathcal{M}$
satisfies $\pi_2[\mathcal{M}] = Z$. Under translation by one
lattice constant, $\mathcal{P}$ transforms as $T_x: \mathcal{P}
\rightarrow - \mathcal{P}$ ($\mathcal{P}$ and $- \mathcal{P}$ both
belong to the same Grassmanian target manifold), and the
coefficient $\Theta$ transforms as $T_x: \Theta \rightarrow -
\Theta$, which guarantees that $\Theta$ is quantized to be
multiple of $\pi$. The same field theory as Eq.~\ref{PL} with a
topological $\Theta$ term has been used to describe the phase
diagram of the integer quantum Hall
systems~\cite{pruisken1,pruisken2,pruisken2011}, while there the
theory is written in the $2d$ real space instead of space-time. A
proposed renormalization group flow for Eq.~\ref{PL} is that,
$\Theta  = 2\pi k$ are stable fixed points, while $\Theta = \pi
(2k+1)$ are instable fixed points which correspond to transitions
between stable fixed points $\Theta = 2k \pi$~\cite{affleck}.

When $\Theta = \pi$, Eq.~\ref{PL} describes the SU($2N$) spin
chain with self-conjugate representation on each site; when
$\Theta = 2\pi$, Eq.~\ref{PL} describes the Haldane phase of a
SU($2N$) spin-chain, or more precisely it is the Haldane phase of
a PSU($2N$) spin chain, as $\mathcal{P}$ is invariant under the
center of SU($2N$). The PSU($2N$) Haldane phase should have
$Z_{2N}$ classification~\cite{psun}, as its boundary could be $2N$
different projective representation of PSU($2N$), which are also
the $2N$ different representation of the $Z_{2N}$ center of
SU($2N$). But the particular state described by Eq.~\ref{P} and
Eq.~\ref{PL} is the ``$N$th" PSU($2N$) Haldane phase, whose $0d$
boundary is a self-conjugate projective representation of
PSU($2N$). This state has a $Z_2$ nature, namely two copies of
this state will be a trivial state, $i.e.$ its boundary is no
longer a projective representation of PSU($2N$).

As we discussed before, the spin-1/2 SU(2) chain can also be
described by Eq.~\ref{o4wzw}, where a VBS order parameter is
introduced. For the SU($2N$) spin chain with self-conjugate
representation, the analogue of Eq.~\ref{o4wzw} is \beqn
\mathcal{S} = \int dx d\tau \ \frac{1}{g} \mathrm{tr}[\partial_\mu
U^\dagger
\partial_\mu U] + \int_0^1 du \ \frac{i2\pi}{ 24 \pi^2}  \mathrm{
tr}[ U^\dagger d U ]^3, \label{U} \eeqn where $U = I_{2N \times
2N}\cos(\theta) + i \sin(\theta) \mathcal{P}$ is a SU($2N$)
unitary matrix. Once again, when $N = 1$, $U$ is a SU(2) matrix,
whose manifold is $S^3$, the same as the target manifold of
Eq.~\ref{o4wzw}. For arbitrary $N$, under translation, $T_x:
\theta \rightarrow \pi - \theta$, $T_x: U \rightarrow - U$. Thus
$\cos(\theta) \sim \phi$ is the VBS order parameter.

The same field theory Eq.~\ref{U} describes the boundary of a $2d$
SPT state with PSU$(2N) \times Z_2$ symmetry, where $Z_2$ plays
the role of $T_x$. And the physical picture of this $2d$ SPT is
that, we decorate every $Z_2$ domain wall with a Haldane phase
with PSU($2N$) symmetry. Thus as one would naively expect, the
SU($2N$) spin chain with self-conjugate representation cannot have
a featureless ground state, because it can be mapped to the
boundary of a nontrivial $2d$ SPT state.

\subsection{SO($N$) spin chain}

A SO($N$) spin chain with a translation symmetry may still obey a
generalization of the LSM theorem. But first let us review the
current understanding of the Haldane phase of $1d$ SO($N$) spin
chain. When $N$ is an odd integer, the double covering group of
SO($N$), $i.e.$ Spin($N$), has a representation which is a spinor
of SO($N$). Thus when $N$ is odd, there is a Haldane phase with
SO($N$) symmetry with a $Z_2$ classification, as two spinors of
SO($N$) will merge into a linear representation of
SO($N$)~\cite{tu08}. Thus in $2d$ space, there is a SPT state with
SO$(N)\times Z_2$ symmetry, which is constructed by decorating the
$1d$ SO($N$) Haldane phase in each $Z_2$ domain wall. Then the
$1d$ boundary of this $2d$ SPT state with SO$(N)\times Z_2$
symmetry, has the feature that, at every $Z_2$ domain wall there
must be a SO($N$) spinor, and this $1d$ boundary cannot be
trivially gapped without breaking the $Z_2$ symmetry.

Now let's consider a Spin($N$) spin chain with a spinor on every
site. Two Spin($N$) spinors with odd $N$ can always form a
singlet, thus this spin chain naturally hosts two fold degenerate
VBS states, which transform into each other through translation by
one lattice constant. The domain wall of these two VBS states is a
Spin($N$) spinor, which is equivalently to the domain wall of the
$Z_2$ order parameter at the $1d$ boundary of the $2d$
SO$(N)\times Z_2$ SPT state mentioned above. Based on these
observations, we can conclude that {\it with odd $N$, a $1d$
Spin($N$) spin chain with spinor representation on every site,
does not permit a featureless gapped state}.

For even $N$, let's take $N = 2n$, then the Haldane phase has a
richer structure. SO($2n$) has a $Z_2$ center which commutes with
all the other elements, thus we can actually consider the Haldane
phase with symmetry PSO($2n$) = SO($2n$)$/Z_2$. Then according to
Ref.~\onlinecite{psun}, the center of Spin($2n$) can be either
$Z_4$ or $Z_2 \times Z_2$, for odd and even integer $n$
respectively. But in either case, a Haldane phase with PSO($2n$)
symmetry could have either spinor or vector representation at the
boundary, both cases are nontrivial Haldane phase. And we can
construct a $2d$ SPT with PSO($2n$)$\times Z_2$ symmetry, by
decorating the $Z_2$ domain wall with a PSO($2n$) Haldane phase.
But this PSO($2n$) Haldane phase must have a $Z_2$ nature, in the
sense that two copies of the Haldane phase must be a trivial
state, because two $Z_2$ domain walls will fuse into a trivial
defect. Thus for both odd and even $n$, we can always decorate the
$Z_2$ domain wall with the PSO($2n$) Haldane phase with a SO($2n$)
vector at the boundary, which leads to the following LSM theorem:

{\it A $1d$ SO($2n$) spin chain with vector representation on
every site does not permit a featureless gapped state. }

This conclusion is consistent with the result of
Ref.~\onlinecite{tulsm}.

\section{spin systems on the square lattice}

\subsection{SU(2) spin systems}

The generalized LSM theorem in higher dimensions does apply to the
$2d$ spin-1/2 system on the square
lattice~\cite{oshikawa,hastings}, $i.e.$ there cannot be a
featureless spin state on the square lattice for a spin-1/2 system
with SU(2) spin symmetry. This conclusion is consistent with many
observations, including a generalization of Eq.~\ref{o4wzw} to
$(2+1)d$~\cite{senthilfisher}: \beqn \mathcal{S} &=& \int d^2x
d\tau \ \frac{1}{g}(\partial_\mu \vec{n})^2 \cr\cr &+& \frac{2\pi
i}{\Omega_4} \int_0^1 du \ \epsilon_{abcde} n^a
\partial_\tau n^b
\partial_x n^c
\partial_y n^d \partial_u n^e, \label{o5wzw} \eeqn where $\vec{n}$
is a five component unit vector, which forms the target manifold
$S^4$ with volume $\Omega_4$. $(n_1, n_2, n_3)$ is still the three
component N\'{e}el order parameter on the square lattice, while
$n_4$ and $n_5$ are the columnar VBS states along the $x$ and $y$
directions respectively. The site-centered 90 degree rotation of
the square lattice acts on $(n_4, n_5)$ as a $Z_4$ rotation, and
close to the deconfined quantum critical
point~\cite{deconfine1,deconfine2}, one can usually embed the
$Z_4$ into an enlarged U(1) group.

The physical meaning of the WZW term in Eq.~\ref{o5wzw} is that,
the vortex of $(n_4, n_5)$ carries a spin-1/2
excitation~\cite{levinsenthil}, and the Skyrmion of $(n_1, n_2,
n_3)$ carries lattice momentum. If we view $b \sim n_4 + in_5$ as
a boson annihilation operator, then the Skyrmion of $(n_1, n_2,
n_3)$ would carry nonzero boson number of $b$. Thus if we destroy
the ordinary N\'{e}el order by condensing the Skyrmions of the
N\'{e}el order parameter, the system automatically develops a
columnar VBS order; and if we destroy the VBS order by condensing
the ($Z_4$) vortex of the columnar VBS order parameter, the system
automatically breaks the spin symmetry and develops the N\'{e}el
order.

Eq.~\ref{o5wzw} can be derived explicitly by starting with the
$\pi-$flux spin liquid state on the square
lattice~\cite{deconfinedual}, and it was proposed as an effective
field theory~\cite{senthilfisher} that describes the deconfined
quantum critical point between N\'{e}el and VBS order on the
square lattice~\cite{deconfine1,deconfine2}, and this is the
critical point whose vicinity we will study and map to the
boundary of a $3d$ system, as we discussed in the introduction.
The key physics of the intertwinement between the N\'{e}el and VBS
order parameter is encoded in the WZW term. Eq.~\ref{o5wzw} is
capable of encapsulating a large SO(5)$\times Z_2^T$ symmetry, and
it also describes the boundary state of a $3d$ bosonic SPT state
whose symmetry can be as large as SO(5)$\times Z_2^T$.
Eq.~\ref{o5wzw} can also describe the boundary of $3d$ SPT states
with a symmetry that is a subgroup of SO(5)$\times
Z_2^T$~\cite{senthilashvin,xuclass}. According to the definition
of SPT states, if the $3d$ bulk is a nontrivial SPT state, then
the boundary cannot be a featureless state; while if the $3d$ bulk
is a trivial direct product state after breaking the SO(5)$\times
Z_2^T$ to its subgroup, then the boundary in principle can be
trivially gapped without degeneracy.

It is clear that if the symmetry SO(5)$\times Z_2^T$ is reduced to
SO(3)$\times$U(1), where $(n_1, n_2, n_3)$ rotates as a vector of
SO(3) and singlet under U(1), while $(n_4, n_5)$ transforms as a
vector of U(1) and singlet of SO(3), the bulk is still a
nontrivial SPT state. And this state can be understood as the
``decorated vortex line" construction introduced in
Ref.~\onlinecite{senthilashvin}: one first breaks the U(1)
symmetry by condensing the two component vector $(n_4, n_5)$, and
decorate a Haldane phase with the SO(3) spin symmetry on each
vortex loop of $(n_4, n_5)$ with odd vorticity, then proliferate
the vortex loops to restore the U(1) symmetry. The SPT state
so-constructed has a $Z_2$ classification, which is consistent
with the $Z_2$ classification of the Haldane phase decorated in
each vortex loop, and also consistent with the $Z_2$ nature of the
fourth Steifel-Whitney class of the SO(5) gauge
bundle~\cite{deconfinedual}. This implies that two copies of the
$3d$ SPT states with SO(3)$\times$U(1) symmetry weakly coupled
together will become a trivial $3d$ bulk state.

The site-centered rotation symmetry of the square lattice acts on
$(n_4, n_5)$ as the $Z_4$ subgroup of U(1). The $3d$ nontrivial
SPT state with SO(3)$\times$U(1) symmetry survives under the
further symmetry breaking of U(1) to $Z_4$, as a $Z_4$ vortex loop
is still a well-define object in the bulk and can be decorated
with a $1d$ Haldane phase. The same conclusion still holds if we
consider a spin-1/2 system on the rectangular lattice. Now this
system corresponds to the boundary of a $3d$ bulk SPT with SO(3)$
\times Z^x_2 \times Z^y_2$. $n_4$, $n_5$ each changes its sign
under one of these two $Z_2$s, while $(n_1, n_2, n_3)$ is odd
under both $Z_2$s. The two $Z_2$s correspond to translation along
$x$ and $y$ directions respectively. The $3d$ bulk SPT state can
be viewed as decorating the $Z^x_2$ domain wall with the $2d$ SPT
with SO(3)$\times Z_2^y$ symmetry, or equivalently decorating the
$Z^y_2$ domain wall with the $2d$ SO(3)$\times Z_2^x$ SPT state.
This observation is consistent with the generalized LSM theorem
which states that a spin-1/2 system on the rectangular lattice
cannot have a featureless state.

Just like the previous section, if we break the spin symmetry down
to $G \rtimes Z_2$, when $G = Z_{2n+1}$ the spin system on the
square lattice allows a featureless state, because the Haldane
phase that we decorated in the vortex loop becomes a trivial state
with only $Z_{2n+1} \rtimes Z_2$ spin symmetry.

Now suppose we consider a spin-1 system on the square lattice,
then a similar deconfined quantum critical point corresponds to
Eq.~\ref{o5wzw} with a level-2 WZW term: the coefficient of the
WZW term doubles. This equation with a level-2 WZW term can be
derived using the $\pi-$flux spin liquid state of a spin-1 system
on the square lattice: there are twice as many Dirac fermions in
the Brillouin zone compared with the case derived in
Ref.~\onlinecite{deconfinedual}, thus the level of the WZW term
also doubles (the difference from the spin-1/2 $\pi-$flux state is
that, the spin-1 $\pi-$flux state has a Sp(4) gauge
fluctuation~\cite{xuspin1}, while the spin-1/2 $\pi-$flux state
has a SU(2) gauge fluctuation). The physical meaning of this term
is that, the vortex of $(n_4, n_5)$ now carries a spin-1 instead
of spin-1/2, which is equivalent to the physics of the boundary of
two weakly coupled $3d$ SPT states with SO(3)$\times$U(1)
symmetry, and as we discussed above, this state is generically a
trivial state in the bulk. Thus its boundary could be a
featureless gapped state. This observation implies that {\it a
spin-1 system on the square lattice permits a featureless state,
which is consistent with the conclusion of
Ref.~\onlinecite{jian1}}.

\subsection{SU($N$) and SO($N$) spin systems}

Now let us consider a SU($N$) spin system on the square lattice,
with fundamental representation (FR) on sublattice A, and
anti-fundamental representation (AFR) on sublattice B. Since the
spins on two nearest neighbor sites can still form a SU($N$) spin
singlet, the columnar VBS order parameter and its $Z_4$ structure
still naturally hold: the site-centered lattice rotation acts as a
$Z_4$ rotation of the columnar VBS order parameter in this system.
The $Z_4$ vortex (antivortex) of the VBS order parameter always
has a vacant sublattice A (B) in the core, hence it always carries
SU($N$) FR (AFR). This is consistent with the fact that a
vortex-antivortex pair can always annihilate, hence the quantum
spin they carry must together form a spin singlet. An analogous
effect on the honeycomb lattice is depicted in
Fig.~\ref{v1},\ref{v2}.

With large enough $N$, a Heisenberg model with the representation
described above should have the four fold degenerate VBS
state~\cite{sachdevread3,sachdevread2}. Now we ask whether a
featureless ground state of this spin system is in principle
allowed or not. Once again, we first view the $Z_4$ lattice
rotation as an onsite internal symmetry, and enlarge it to U(1).
Then the $2d$ spin system on the square lattice can be {\it
potentially} viewed as the boundary of a $3d$ bosonic SPT state
with PSU$(N)\times U(1)$ symmetry.

The bosonic SPT states with PSU$(N)\times U(1)$ symmetry do exist
in $3d$, and they can be interpreted as the decorated vortex loop
construction, $i.e.$ we decorate every U(1) unit vortex loop with
a $1d$ PSU($N$) Haldane phase, whose boundary is a projective
representation of the PSU($N$), or a faithful representation of
SU($N$). As we have discussed, $1d$ PSU($N$) Haldane phase has a
$Z_N$ classification, which corresponds to $N$ different
projective representations of the PSU($N$) group, or $N$ different
representation of the $Z_N$ center of SU($N$).

In general, the $N-1$ different nontrivial Haldane phases of
PSU($N$) can be described by Eq.~\ref{PL} with $\Theta = 2\pi$,
and $\mathcal{P}$ replaced by~\cite{affleck} \beqn \mathcal{P} = V
\Omega V^\dagger, \;\;\Omega \equiv\! \left(
\begin{array}{cccc}
\mathbf{1}_{m \times m} & \mathbf{0}_{ m \times N-m} \\ \\
\mathbf{0}_{N-m \times m} & \;\;\mathbf{-1}_{N-m \times N-m}
\end{array}
\right) \;\label{PP}\eeqn with $m = 1, \cdots N-1$, and $V$ is a
SU($N$) matrix. All the configurations of $\mathcal{P}$ belong to
the Grassmanian manifold $U(N)/[U(m) \times U(N-m)]$. In our case,
when the vortex line terminates at the boundary, the vortex at the
boundary will carry a FR of SU($N$), hence for our case we need to
choose $m = 1$, and $\mathcal{P}$ becomes the CP$^{N-1}$ manifold.

%Formally, a $3d$ SPT with PSU$(N)\times U(1)$ symmetry can be
%described as following: we can first embed the CP$^{N-1}$ manifold
%target space of the PSU($N$) order parameter and a U(1) vector
%(more precisely a SO(2) vector) $(n_1, n_2)$ with unit length into
%a larger manifold $\mathcal{M} = U(2N)/[U(N)\times U(N)]$, by
%defining a larger order parameter $\tilde{\mathcal{P}}$: \beqn
%\tilde{\mathcal{P}} = \mathcal{P} \otimes \tau^z \cos(\theta)  +
%I_{N\times N} \otimes (n_1 \tau^x + n_2 \tau^y) \sin(\theta).
%\eeqn Then because of the fact $\pi_4[\frac{U(2N)}{U(N)\times
%U(N)}] = Z$ for $N \geq 2$, the following topological
%$\Theta-$term can be defined for $\tilde{\mathcal{P}}$ in the $3d$
%bulk: \beqn \int d^3x d\tau \ \frac{i
%2\pi}{256\pi^2}\mathrm{tr}[\tilde{\mathcal{P}}
%\partial_\mu \tilde{\mathcal{P}} \partial_\nu \tilde{\mathcal{P}} \partial_\rho
%\tilde{\mathcal{P}}
%\partial_\lambda \tilde{\mathcal{P}}]\epsilon_{\mu\nu\rho\lambda}.
%\eeqn When expanded with $\mathcal{P}$ and $(n_1, n_2)$, this
%topological term precisely captures the physics of decorating each
%vortex of $(n_1, n_2)$ with a PSU($N$) Haldane phase.

However, let us not forget that eventually we need to break the
U(1) symmetry down to $Z_4$. Then for the $3d$ SPT state to
survive under this symmetry breaking, the $Z_N$ classification of
the PSU($N$) Haldane phase must be compatible with the $Z_4$
vortex. If $N$ and $4$ are coprime, then this bulk state
definitely becomes trivial after breaking the U(1) to $Z_4$. For
example, when $N = 3$, there is no consistent way we can decorate
the $Z_4$ vortex with a PSU(3) Haldane phase. Because four $Z_4$
vortex loops merge together will no longer be a well-defined
defect, while four PSU(3) Haldane phases merge together is still a
nontrivial Haldane phase. Thus for odd integer $N$, the $3d$ SPT
phase with PSU$(N) \times$U(1) symmetry becomes a trivial phase
once U(1) is broken down to $Z_4$.

To further demonstrate that for odd integer $N$, the $3d$ SPT
phase with PSU$(N) \times$ U(1) symmetry is trivialized with U(1)
broken down to $Z_4$, we need to show that its $2d$ boundary can
be trivially gapped out when U(1) is broken down to $Z_4$. One of
the $2d$ boundary states of the $3d$ PSU$(N) \times$U(1) SPT
phase, is a $Z_N$ topological order, which can be constructed by
starting with a superfluid order with spontaneous U(1) symmetry
breaking at the $2d$ boundary, and then condense the $N-$fold
vortex (a vortex with $2\pi N$ vorticity) of the superfluid order.
The single vortex of the superfuid phase carries a FR of SU($N$),
hence a $N-$fold vortex can carry a SU($N$) singlet, and its
condensate is a $Z_N$ topological order which preserves all the
symmetries. A $2d$ $Z_N$ topological order has bosonic $e$ and $m$
excitations, while $e$ and $m$ have mutual statistics with
statistical angle $\theta_{e,m} = 2\pi/N$. In our construction,
the $e$ excitation carries $1/N$ charge of the U(1) symmetry, and
the $m$ excitation carries a FR of SU($N$).

Once U(1) is broken down to $Z_4$,
%a four-body bound state of $e$
%will carry charge $4/N$ under the $Z_4$ group, and it acquires
%phase angle $2\pi/N$ under a unit $Z_4$ rotation. This $Z_4$
%rotation can be cancelled by a $Z_N$ gauge transformation. Now
%suppose we condense a bound state of four $e$ particles, the $Z_N$
%topological order is driven into a completely featureless gapped
%state without any anyons, and all the global symmetries are
%preserved. This is only possible when $N$ is an odd integer.
In order to gap out the $Z_N$ topological order, we can condense
the bound state of a $e$ particle and $3N$ $Z_4$ charges. This
bound state carries $\frac{3N^2 + 1}{N}$ $Z_4$ charges. Under the
$Z_4$ transformation, it acquires a phase $\exp\left(
\frac{2\pi(3N^2+1)}{4N} i \right)$, which can always be
cancelled/compensated by a gauge transformation with odd integer
$N$ (the numerator of the phase angle is always a multiple of
$8\pi$ with odd integer $N$). Thus the condensate of this bound
state will drive the $Z_N$ topological order into a completely
featureless gapped state without any anyons, and all the global
symmetries are preserved. This is only possible when $N$ is odd.

As a contrast, for even integer $N$, we can always construct a
nontrivial $3d$ SPT by decorating the $Z_4$ vortex loop with the
$1d$ SPT state with SU$(N)/Z_2$ symmetry, which has a $Z_2$
classification.

Now we can make the following conclusion:

{\it A SU($N$) spin system on the square lattice with fundamental
and anti-fundamental representation on the two sublattices, permit
a featureless gapped ground state for odd integer $N$.}

We can also consider SO($N$) spin systems on the square lattice.
The analysis is very similar to the previous case. We can make the
following conclusion:

{\it A SO($2n$) spin system with vector representation on every
site does not permit a featureless gapped state on the square
lattice. }

{\it A SO($2n+1$) spin system with spinor representation on every
site does not permit a featureless gapped state on the square
lattice. }

On the other hand, A SO($2n+1$) spin system with vector
representation on every site does permit a featureless gapped
state.

\section{spin systems on the honeycomb lattice}

\subsection{SU(2) spin systems}

A spin-1/2 system on the honeycomb lattice, when tuned close to
certain point, can also be described by Eq.~\ref{o5wzw}.
Eq.~\ref{o5wzw} can be derived with the SU(2) spin liquid on the
honeycomb lattice, like the one discussed in
Ref.~\onlinecite{hermele2007}. Now the lattice symmetry, both the
translation $T_x$ and a site-centered 120 degree rotation, acts as
a $Z_3$ subgroup of the U(1) transformation on $(n_4, n_5)$.

\begin{figure}[tbp]
\begin{center}
\includegraphics[width=200pt]{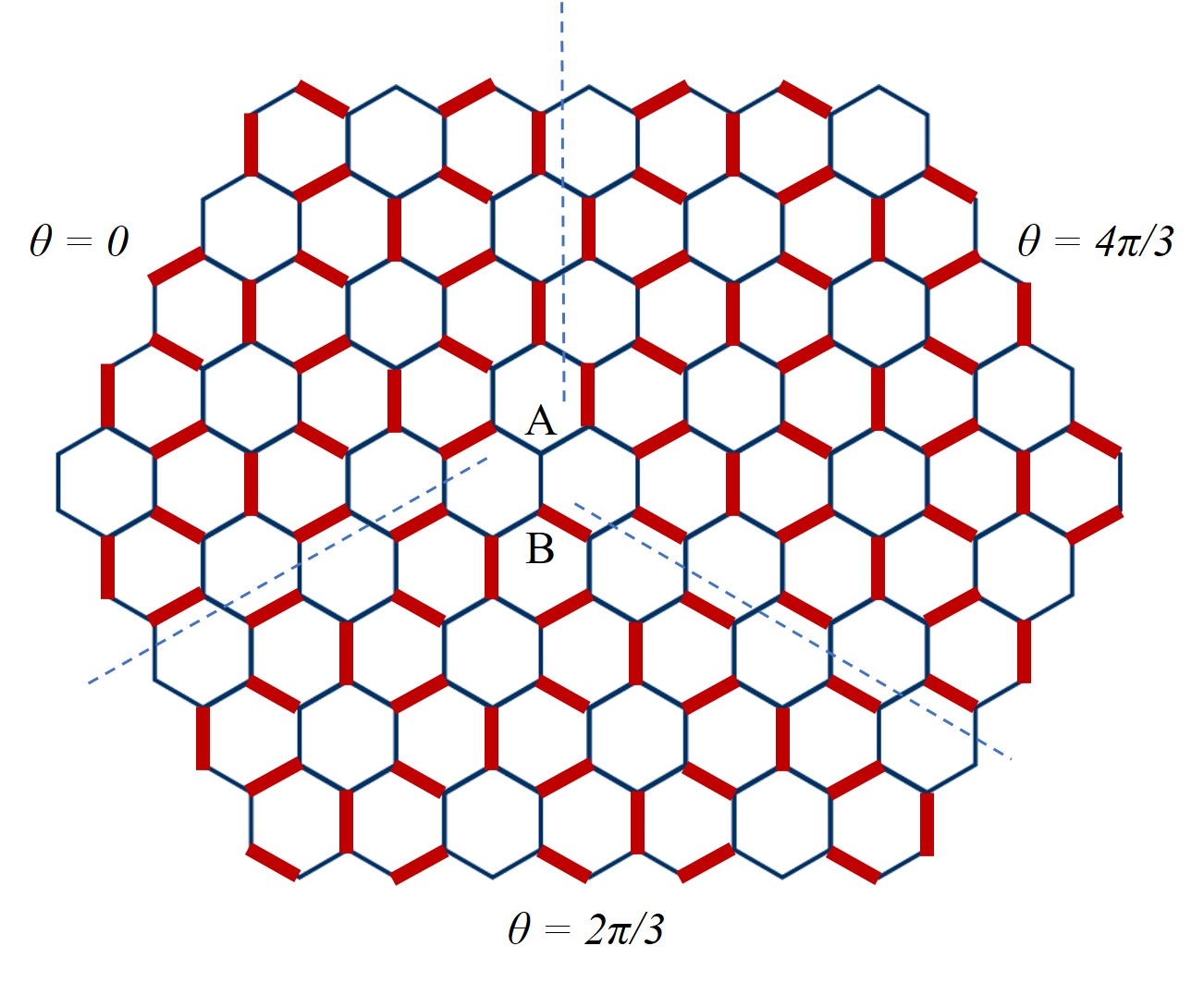}
\caption{A $Z_3$ vortex of the VBS order parameter on the
honeycomb lattice has a vacant site on the sublattice A, and hence
carries a fundamental representation of the SU($N$) spin.}
\label{v1}
\end{center}
\end{figure}

\begin{figure}[tbp]
\begin{center}
\includegraphics[width=200pt]{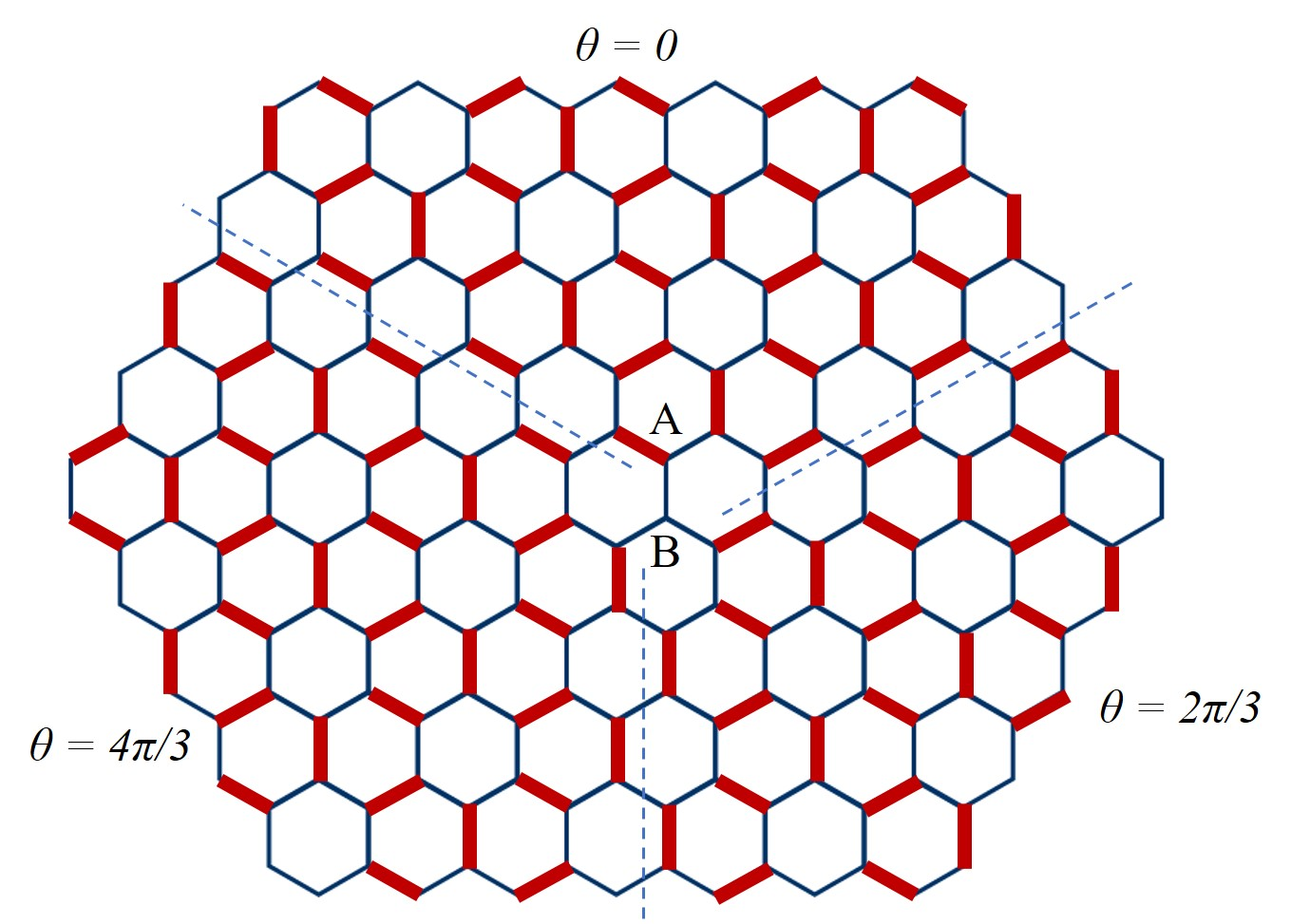}
\caption{An antivortex of the VBS order parameter on the honeycomb
lattice has a vacant site on the sublattice B, and hence carries a
anti-fundamental representation of the SU($N$) spin.} \label{v2}
\end{center}
\end{figure}

Once again, the question of whether a featureless spin-1/2 state
exists on the honeycomb lattice is equivalent to whether the $3d$
SPT state with SO(3)$\times$U(1) symmetry is stable against
breaking the U(1) down to $Z_3$. It turns out that this time the
$3d$ bulk becomes a trivial state. The vortex loop decoration
picture fails with a $Z_3$ symmetry. Suppose we decorate a Haldane
phase on each $Z_3$ vortex loop, then three of the $Z_3$ vortex
loops would be decorated with three Haldane phases, and due to the
$Z_2$ classification of the $1d$ Haldane phase, three Haldane
phases is still a nontrivial $1d$ SPT state. However, a three fold
$Z_3$ vortex loop is no longer a well-defined defect any more.
Thus the decorated vortex loop picture is incompatible with the
$Z_3$ symmetry. Thus the bulk becomes a trivial state once we
break the U(1) down to $Z_3$. This implies that the $2d$ boundary,
which corresponds to the spin-1/2 system on the honeycomb lattice,
permits a featureless spin state. This is consistent with the
previous result on the honeycomb lattice~\cite{jian1,jian2}.

We can also add other symmetries of the honeycomb lattice, such as
reflection $P_x: y \rightarrow -y$. Under this reflection, $ P_x:
(n_1, n_2, n_3) \rightarrow - (n_1, n_2, n_3)$, while $(n_4, n_5)$
is unchanged. In the Euclidean space-time, a reflection symmetry
can be treated equivalently as the time-reversal symmetry.  Thus
with both translation $T_x$ and reflection $P_x$, we need to study
whether the $3d$ SPT state with SO(3)$\times Z_2^T \times $U(1)
symmetry is stable against symmetry breaking down to SO(3)$\times
Z_2^T \times Z_3$. The analysis is the same as before: the $3d$
SPT state with SO(3)$\times Z_2^T \times $U(1) symmetry is
constructed with proliferated vortex loops decorated with a $1d$
Haldane phase with SO(3)$\times Z_2^T$ symmetry. However, this
construction is still incompatible with the $Z_3$ vortex loops,
because the classification of the Haldane phase with SO(3)$\times
Z_2^T$ symmetry is $Z_2\times Z_2$.

\subsection{SU($N$) and SO($N$) spin systems}

Now let us consider a SU($N$) spin system on the honeycomb
lattice, again with FR on sublattice A, and AFR on sublattice B.
This system can still form the three fold degenerate VBS states,
and the vortex (antivortex) of the VBS order parameter has a
vacant site in sublattice A (B), which carries a FR (AFR) of
SU($N$) (Fig.~\ref{v1},\ref{v2}).

Now we want to ask whether the $3d$ SPT state with PSU($N$)$\times
U(1)$ symmetry is stable against breaking the U(1) down to $Z_3$.
This depends on whether the PSU($N$) SPT state decorated on the
vortex line is compatible with the $Z_3$ nature of the vortex
line, $i.e.$ $N$ at least cannot be coprime with $3$. Thus when
$N$ is coprime with $3$, the $3d$ SPT state PSU($N$)$\times U(1)$
symmetry is trivialized by breaking U(1) down to $Z_3$.

Just like the case in the previous section, the boundary of a $3d$
SPT with PSU($N$)$\times U(1)$ symmetry could be a $2d$ $Z_N$
topological order, whose $e$ particle carries $1/N$ charge of
U(1), and $m$ particle carries a FR of SU($N$). Once U(1) is
broken down to $Z_3$, if $N$ is coprime with 3, by condensing a
bound state of $e$ and certain number of $Z_3$ charges, this $2d$
boundary $Z_N$ topological order is driven into a featureless
gapped state.

We can now make the following conclusion:

{\it SU($N$) spin systems on the honeycomb lattice with
fundamental and anti-fundamental representation on the two
sublattices, permit a featureless gapped ground state when $N$ is
not a multiple of 3.}

Also, similar conclusions can be made for SO($N$) spin systems:

{\it A SO($2n$) spin system with vector representation on every
site permits a featureless state on the honeycomb lattice. }

{\it A SO($2n+1$) spin system with spinor or vector representation
on every site also permits a featureless state on the honeycomb
lattice.}

\section{$3d$ spin systems on the cubic lattice}

A spin-1/2 system on the cubic lattice is subject to the
generalized LSM theorem, thus it cannot have a featureless state.
Besides the common N\'{e}el ordered state, another natural
spin-1/2 state on the cubic lattice is the columnar VBS state. And
the ``hedgehog monopole" of the VBS order parameter carries a
spin-1/2, and the monopole of the N\'{e}el order parameter carries
lattice momentum~\cite{senthilloop}, whose condensate is precisely
the VBS order. This system enjoys a nice self-duality structure.
We can introduce the vector N\'{e}el order parameter $\vec{n}^{e}$
and vector VBS order parameter $\vec{n}^{m}$, as well as their
CP$^1$ fields~\cite{senthilloop,chenbalents}: \beqn \vec{n}^{e}
\sim \frac{1}{2}z^{e \dagger}\vec{\sigma} z^e, \ \ \ \vec{n}^{m}
\sim \frac{1}{2}z^{m \dagger}\vec{\sigma} z^m. \eeqn When the spin
system is driven into a photon phase, which is stable in $(3+1)d$,
$z^e$ and $z^m$ are the gauge charge and the Dirac monopole of the
dynamical U(1) gauge field $a_\mu$ respectively.

The cubic lattice symmetry acts on $\vec{n}^{m}$ as the octahedral
subgroup of SO(3)~\footnote{The octahedral group O does not
include the spatial mirror (reflection) symmetry. The mirror
symmetry is equivalent to time-reversal symmetry in the analysis
of SPT states, as we explained previously. Including the mirror
symmetry does not change our conclusions, because the SO(3)
Haldane phase with or without an extra time-reversal symmetry
always has a $Z_2$ nature, $i.e.$ two of these Haldane phases
coupled together becomes a trivial state.}, and $z^e$, $z^m$ carry
projective representation of the SO(3) spin and (enlarged) SO(3)
lattice symmetry respectively. The intertwinement between the
N\'{e}el and VBS order is captured by a $(3+1)d$ WZW term of a six
component vector which contains both $\vec{n}^e$ and
$\vec{n}^m$~\cite{ryuwzw}.

The same physics can be realized at the boundary of a $4d$ SPT
state with SO(3)$^e \times $SO(3)$^m$ symmetry. This state can be
understood as the ``decorated monopole line" construction. In the
$4d$ space, a SO(3)$^e$ hedgehog monopole is a line defect, and we
can decorate it with a $1d$ Haldane phase with SO(3)$^m$ symmetry.
The CP$^1$ field $z^m$ can be viewed as the termination of the
SO(3)$^e$ hedgehog monopole line at the $3d$ boundary, which is
also the boundary state of the $1d$ SO(3)$^m$ Haldane phase. The
self-duality of the boundary QED implies that the decoration
construction is necessarily mutual, $i.e.$ we must simultaneously
decorate the SO(3)$^m$ hedgehog monopole with a Haldane phase with
the SO(3)$^e$ symmetry.

The ``mutual decoration" construction can also be perceived as
follows. In the $4d$ space, we can discuss the braiding process of
two loops. Imagine we create two loops $L^e$ and $L^m$ from
vacuum, and annihilate them at a later time, then the world sheets
of both loops are topologically two dimensional spheres, labelled
as $S^2_e$ and $S^2_m$. If these two loops are braided, their
world sheets are linked in the five dimensional space-time. This
linking can be interpreted as the intersection of $S^2_e$ with the
interior of $S^2_m$ (which is a three dimensional ball) at one
point in the space time. Now suppose $S^2_e$ and $S^2_m$ are the
world sheets of the SO(3)$^e$ and SO(3)$^m$ monopole lines
respectively, if the SO(3)$^m$ monopole line is decorated with the
SO(3)$^e$ Haldane phase, then this linking will accumulate phase
$2\pi$, which comes from the $\Theta-$term of the SO(3)$^e$
Haldane phase.

The linking mentioned above is also symmetric under interchanging
$e$ and $m$, namely it can be viewed as the intersection of
$S^2_m$ with the interior of $S^2_e$ at another point in the
space-time. Thus if this linking accumulates phase $2\pi$, then
consistency demands that the SO(3)$^e$ monopole line be decorated
with the SO(3)$^m$ Haldane phase too.

%and it can also be interpreted as the intersection of $S^2_m$ with
%the interior of $S^2_e$ ($D^3_e$) at another point.

%Topologically, the $4d$ space with a SO(3) monopole loop is
%equivalent to $S^2 \times S^1$. We can create a SO(3)$^e$ and a
%SO(3)$^m$ monopole loop from vacuum, then sweep the SO(3)$^e$
%monopole loop along the $S^2$ around the SO(3)$^m$ monopole loop,
%then annihilate both loops. Topologically this creates a link
%between the world sheets of both monopole loops (both world sheets
%are $S^2$) in the five dimensional space-time, and this process is
%symmetric under exchanging $e$ and $m$. If this process
%accumulates phase $2\pi$ for the partition function, it can be
%viewed either as the contribution from the SO(3)$^e$ Haldane phase
%decorated on the SO(3)$^m$ monopole loop, or vice versa.

The $4d$ SPT state so constructed obviously has a $Z_2$
classification, as both the SO(3)$^e$ and SO(3)$^m$ SPT phases
have $Z_2$ classification. To make an explicit connection with the
$(3+1)d$ QED state discussed in Ref.~\onlinecite{senthilloop}, one
can first start with fractionalizing $\vec{n}^e$ in the bulk, and
introduce a $(4+1)d$ U(1) gauge field $a_\mu$. The hedgehog
monopole line of $\vec{n}^{e}$ becomes the Dirac monopole line of
$a_\mu$, which is decorated with the SO(3)$^m$ Haldane phase. Now
we condense the Dirac monopole line in the bulk, but do not
condense the termination of the Dirac monopole line at the $3d$
boundary, which becomes the Dirac monopoles (point like defects)
at the $3d$ boundary. This will lead to a gapped $4d$ bulk state,
while the $3d$ boundary is the QED state discussed in
Ref.~\onlinecite{senthilloop} with $z^e$ and $z^m$ being the gauge
charge and Dirac monopole respectively.

The picture above can again be generalized to the PSU($N$) spin
system with FR and AFR on the two sublattices. Whether this spin
system permits a featureless gapped state or not, is equivalent to
whether the corresponding $4d$ bulk state is a trivial state or a
SPT state. The CP$^{N-1}$ manifold, $i.e.$ the SU($N$)
generalization of the N\'{e}el order parameter, has
$\pi_2[\mathrm{CP}^{N-1}] = Z$, and hence also has a ``hedgehog
monopole" line in the $4d$ space. Thus we can again decorate the
SO(3)$^m$ monopole line with the PSU($N$) Haldane phase, and
simultaneously decorate the PSU($N$) monopole line with the
SO(3)$^m$ Haldane phase. But now this $4d$ state is {\it not}
always a nontrivial SPT state. Because the SO(3)$^m$ Haldane phase
has a $Z_2$ classification, hence even-number copies of the $4d$
state must be a trivial state, while odd-number copies of the
states is equivalent to the state itself. On the other hand, the
PSU($N$) Haldane phase has a $Z_N$ classification, namely $N$
copies of the states must be trivial. Thus the $4d$ bulk state so
constructed has a $Z_{(2,N)}$ classification: the ``mutual
monopole line decoration" gives us a nontrivial $4d$ SPT state
only with even $N$.

The natural $3d$ boundary state of the $4d$ bulk based on the
``mutual" monopole line decoration construction, is a U(1) photon
phase whose $e$ excitations carry SU($N$) fundamental, and $m$
carries a spin-1/2 of SO(3). When $N$ is odd, we can drive the
$3d$ boundary into a featureless state by condensing the dyon
which is a bound state of $N$ $e$ particles and two $m$ particles.
We label this dyon as the $(N, 2)$ dyon. This $(N,2)$ dyon is a
boson, and its condensate will gap out the photons, while
confining all the point particles, because there is no point
particle that is mutual bosonic with this dyon, except for the
dyon itself. Also, the $(N, 2)$ dyon could be a singlet of
SU($N$), and singlet of SO(3), thus its condensate does not break
any global symmetry. This means that for odd integer $N$, the $3d$
boundary of the $4d$ bulk state can be driven into a featureless
gapped state, which again demonstrates that the $4d$ bulk state
constructed above is trivial when $N$ is odd.

By contrast, if $N$ is even, then the $(N/2, 1)$ dyon (with
nontrivial representation of SU($N$) and SO(3)) is still
deconfined in the condensate of $(N, 2)$ dyon, and this condensate
has topological order.

Now we can conclude that:

{\it For odd $N$, the SU($N$) spin system on the cubic lattice
with FR and AFR spins on two sublattices permits a featureless
spin state.}

Here we propose a low energy effective field theory for the $4d$
SPT state that captures the ``mutual decorated monopole line"
construction. We first define a U($2N$) matrix field $U$ as \beqn
U = \cos(\theta) \mathcal{P} \otimes I_{2\times 2} + i
\sin(\theta) I_{N\times N} \otimes \vec{n} \cdot \vec{\tau}, \eeqn
where $\mathcal{P}$ is the CP$^{N-1}$ matrix field given by
Eq.~\ref{PP}. The ``mutual decoration" picture is captured by a
topological term in the nonlinear sigma model of $U$ which reads
\beqn \label{5dtheta} \mathcal{L}^{topo}_{5d} = \int d^4x d\tau \
\frac{2\pi }{480 \pi^3} \Tr\left[(U^\dagger dU)^5\right]. \eeqn

We will show that if we manually create a monopole line of
$\vec{n}$, the topological term Eq.~\ref{5dtheta} precisely
reduces to the topological term of the $(1+1)d$ PSU$(N)$ SPT. Let
us parametrize the $(4+1)d$ space-time by Cartesian coordinates
$(x,y,z,w,\tau)$ and consider a static monopole line of $\vec{n}$
whose core line lies on the $w$-axis. For any fixed $w$ and
$\tau$, we will see a monopole configuration of $\vec{n}$ centered
at origin in the $xyz$ space. For a monopole configuration in the
$xyz$ space, we have \beqn &&\theta(r=0)=0, \cr \cr &&
\theta(r\rightarrow\infty)=\pi/2 \cr\cr &&\int_{r=r_0>0} \
d^2\Omega \
\frac{1}{8\pi}\epsilon_{ijk}\epsilon_{\alpha\beta}n^i\partial_{\alpha}n^j\partial_{\beta}n^k=1
\eeqn where $r=\sqrt{x^2+y^2+z^2}$. We also assume $\mathcal{P}$
is a function of $w$ and $\tau$. Now we plug in this configuration
of $\vec{n}$ in to Eq.~\ref{5dtheta} and integrate over $x,y$ and
$z$ directions. This topological term reduces to the following
$(1+1)d$ topological term in the $(w,\tau)$ space: \beqn
\mathcal{L}^{topo}_{2d} = \int dwd\tau \ \frac{2\pi
}{16\pi}\epsilon_{\mu\nu}\Tr\left(\mathcal{P}\partial_{\mu}\mathcal{P}\partial_{\nu}\mathcal{P}\right),\eeqn
which is precisely the topological $\Theta$-term for the PSU$(N)$
Haldane phase. This indicates that Eq.~\ref{5dtheta} implies there
is a PSU$(N)$ SPT decorated on the monopole line of $\vec{n}$.

If we consider a monopole line of $\mathcal{P}$ along $w$-axis,
then in the $xyz$ directions we have \beqn &&\theta(r=0)=\pi/2,
\cr \cr && \theta(r\rightarrow\infty)=0 \cr\cr &&\int_{r=r_0>0} \
d^2\Omega \
\frac{i}{16\pi}\epsilon_{\mu\nu}\Tr\left(\mathcal{P}\partial_{\mu}\mathcal{P}\partial_{\nu}\mathcal{P}\right)=1
\eeqn Now integrating over $x, y$ and $z$ directions will give us
the following topological term in the $(1+1)d$ space-time of the
monopole line world sheet: \beq \mathcal{L}^{topo}_{2d} = \int
dwd\tau \ \frac{2\pi
i}{8\pi}\epsilon_{abc}\epsilon_{\mu\nu}n^a\partial_{\mu}n^b\partial_{\nu}n^c,
\eeq which exactly corresponds to the topological term of the
$(1+1)d$ SO$(3)$ Haldane phase. Therefore, the topological term in
Eq.~\ref{5dtheta} captures the ``mutual decoration" construction
of the $(4+1)d$ SPT phase with PSU$(N)\times$SO$(3)$ symmetry.

\section{Further proof of our conclusions}

\subsection{Explicit construction of featureless spin states}

Let us first restate our main conclusions about SU($N$) spin
systems on the square, honeycomb, and cubic lattices:

{\it 1. A SU($N$) spin system on the square lattice with
fundamental (FR) and anti-fundamental representation (AFR) on the
two different sublattices respectively, permits a featureless
gapped ground state when $N$ is an odd integer;}

$ $

{\it 2. A SU($N$) spin system on the honeycomb lattice with FR and
AFR on two different sublattices, permits a featureless gapped
ground state when $N$ is coprime with 3.}

$ $

{\it 3. A SU($N$) spin system on the cubic lattice with FR and AFR
spins on two different sublattices, permits a featureless spin
state when $N$ is odd.}

$ $

For all the spin systems listed above, we can construct explicit
featureless tensor product spin states similar to the AKLT states.
All these states will be discussed in a future
work~\cite{jianfuture}. Here we discuss some of the examples of
this construction.

On the honeycomb lattice, in the case of $N = 3k+1$, we introduce
$3k$ auxiliary spins on each site. We also introduce a tensor on
each site: \beqn T^{\alpha}_{i_1 i_2 \cdots i_{3k}} =
\varepsilon_{\alpha i_1 i_2 ... i_{3k}},
\label{Eq:Honeycomb_3k+1_Proj}\eeqn where $\varepsilon_{\alpha i_1
i_2 ... i_{3k}}$ is the total anti-symmetric tensor with $N=3k+1$
indices. Here, the $i_1, i_2, \cdots i_{3k}$ labels the $3k$
auxiliary FR (or AFR) spin degrees of freedom on each site in
sublattice B (or A) before the projection. Each label $i_n$ takes
value in $1,2, \cdots N$ representing the $N$ states in each FR
(or AFR). The label $\alpha$, which also takes value $1,2, \cdots
N$, represents the physical states in AFR (or FR) spin degrees of
freedom on each site in sublattice B (or A). Physically, on each
site of sublattice A, the tensor in
Eq.~\ref{Eq:Honeycomb_3k+1_Proj} projects the $3k$ auxiliary AFR
spins into a totally anti-symmetric channel which, due to the
nature of $\SU(3k+1)$, becomes the physical FR spin. The analysis
for sites in the sublattice B is similar. Now we can use the
auxiliary spins to construct a featureless gapped state on the
honeycomb lattice with $k$ SU$(N)$ singlet bonds along each link
of the lattice, which is reminiscent of the AKLT state.

%Here, since we eventually project all the spin on each site into a
%totally anti-symmetric channel, there is no need to specify which
%auxiliary spin each label $i_n$ refers to before the projection.

Obviously, the so constructed tensor product state respects the
translation symmetry of the lattice. Now we analyze the
compatibility between the point group $C_{3v}$ and the site tensor
in Eq.~\ref{Eq:Honeycomb_3k+1_Proj}. Here, notice that we include
not only the $C_3$ rotation symmetry but also the mirror
reflection symmetry of the honeycomb lattice into consideration.
We notice that the point group only induces a permutation of the
singlet bonds before the projection. Therefore, the action of the
point group permutes the $3k$ spins on each site. Since we project
the $3k$ spins into a totally anti-symmetric channel using the
site tensor, the point group induced permutation keeps the site
tensor invariant up to a global sign which is unimportant for the
global tensor network wave function. Therefore, we can conclude
that the choice of projection tensor in
Eq.~\ref{Eq:Honeycomb_3k+1_Proj} preserves the space symmetries.

On the square lattice, in the case of $N = 4k+1$, we introduce
$4k$ auxiliary spins on each site and let the auxiliary spins form
a state with $k$ SU$(N)$ singlet bonds along each link of the
square lattice. We can choose the site tensors to be \beqn
T^{\alpha}_{i_1 i_2 \cdots i_{4k}} = \varepsilon_{\alpha i_1 i_2
\cdots i_{4k}}, \label{Eq:Square_4k+1_Proj} \eeqn where
$\varepsilon_{\alpha i_1 i_2 \cdots i_{4k}}$ is the total
anti-symmetric tensor with $N=4k+1$ indices. Based on analysis
completely in parallel with the honeycomb lattice, we conclude
that the physical spin carries AFR (FR) under $\SU(N)$ if the
auxiliary spins transform as FR (AFR). Also, we can conclude that
the tensors in Eq. \ref{Eq:Square_4k+1_Proj} are invariant under
the $C_{4v}$ point group action up to an unimportant sign because
the actions of the $C_{4v}$ point group on the site tensor are
only permutation of the tensor indices. Now we can use the $4k$
auxiliary spins on each site to construct a featureless spin state
on the square lattice.

\begin{figure}[tbp]
\begin{center}
\includegraphics[width=220pt]{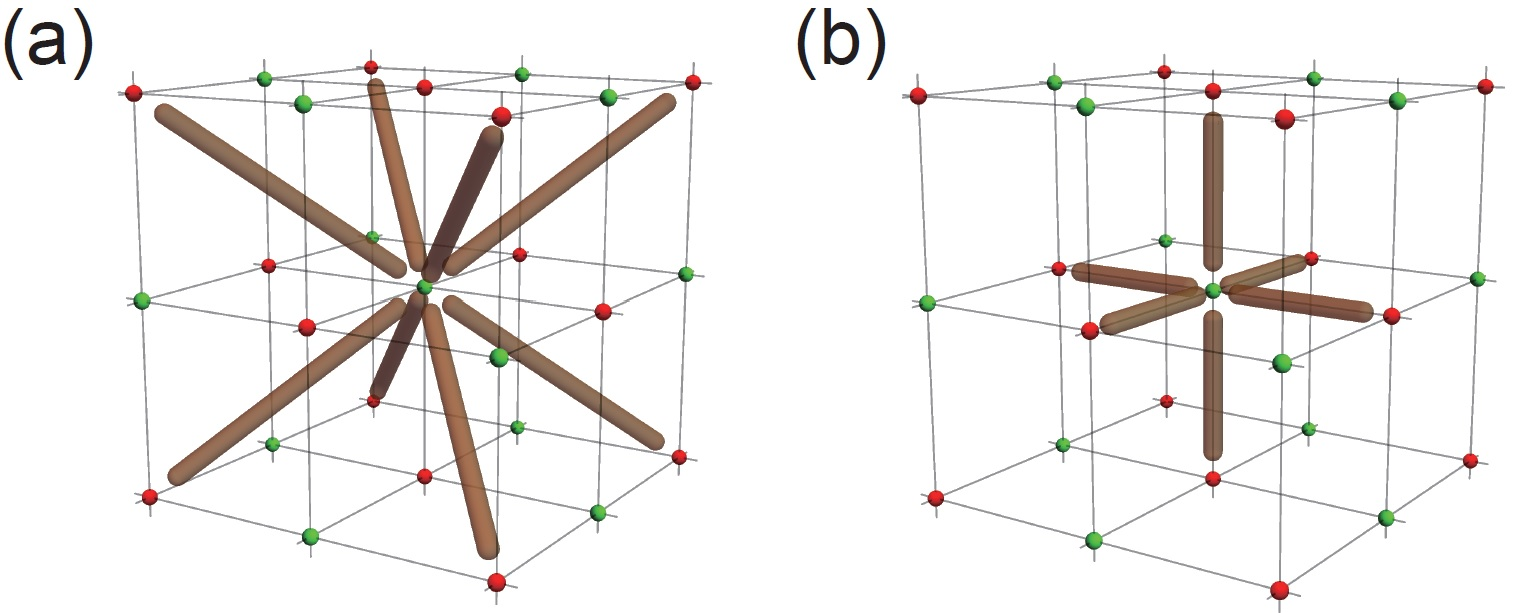}
\caption{$(a)$ The schematic featureless SU($N$) spin state on the
cubic lattice when $N = 8p +1$; $(b)$, the schematic featureless
SU($N$) spin state on the cubic lattice when $N = 6q + 1$. More
general spin systems with $N = 8 p + 6 q + 1$ have valence bonds
extended along both the link and diagonal directions of the cubic
lattice.} \label{cubic}
\end{center}
\end{figure}

On the cubic lattice, for any odd integer $N$ that is not $3,5$ or
$11$, we can write $N$ as $N=8 p + 6 q + 1$ with $p$ and $q$
non-negative integers.
%In this case, the construction of AKLT
%states starts with $k$ copies of $|\VBScbO \rangle$ and $q$ copies
%of $|\VBScbH \rangle$ which are, as a whole, defined on the
%$(N-1,N-1)$-cubic lattice. On a single site of the
%$(N-1,N-1)$-cubic lattice, we use $i_n=1,2...,N$ to label the
%states associated the $n$th auxiliary spin on that site ($1\leq
%n\leq N-1$).
Again we introduce $N-1$ auxiliary spins, and an on-site tensor
$T^{\alpha}_{i_1 i_2 ....i_{N-1}} = \varepsilon_{\alpha i_1 i_2
... i_{N-1}}$. Namely on sublattice B, we represent the AFR with
$N-1$ auxiliary FRs, and on sublattice A we represent the FR with
$N-1$ AFRs. Now these auxiliary spins can form a featureless
states with valence bonds extended either along the link (for $N =
6q + 1$) or the diagonal directions (for $N = 8p + 1$), or both
directions (when $p$ and $q$ are both nonzero) on the cubic
lattice (Fig.~\ref{cubic}).

The point group $O_h$ of the cubic lattice will induce a
permutation among the $N-1$ auxiliary spins on each site which at
most leads to an unimportant sign change of the site tensor.
Therefore, this site tensor is compatible with the point group
$O_h$ symmetry. In fact, the $O_h$ point group is isomorphic to
$S_4 \times Z_2$. The $Z_2$ part is the spatial inversion which
takes the point $(x,y,z)$ to $(-x,-y,-z)$. $S_4$ is the
permutation group of 4 elements, which can be generated by a $Z_3$
cyclic permutation and a $Z_4$ cyclic permutation. In the language
of the point group, the $S_4$ part is the part of $O_h$ that
preserves the spatial orientation. It can be generated by a $C_3$
rotation about the $(1,1,1)$-axis and a $C_4$ rotation about the
$z$-axis. This $S_4$ part alone (without the spatial inversion) is
usually referred to the point group $O$.

The construction of these featureless tensor product wave
functions does provide strong evidence to our conclusions in
previous sections. Nevertheless, we need to comment that, to
eventually confirm the featureless-ness of these tensor product
wave functions, numerical simulation of these states is demanded,
in order to rule out possible spontaneous symmetry breaking, etc.
For instance, it is known that the AKLT wave function on a three
dimensional lattice could have long range spin order.

\subsection{Connection to ``lattice homotogy class"}

In fact, we can also simplify all the discussions by just
considering a $Z_N \times Z_N$ subgroup of $\PSU(N)$ and analyzing
how the FR and AFR of $\SU(N)$ transform under this $Z_N \times
Z_N$ subgroup. To specify this $Z_N \times Z_N$ subgroup, we first
consider two $\SU(N)$ matrices in the FR: \beqn && g_1 = e^{i \pi
(N-1)} \left(
\begin{array}{ccccccc}
0 & 1 & 0 & ... & 0 & 0 \\
0 & 0 & 1 & \ddots &  & 0  \\
0 & 0 & \ddots & \ddots &  \ddots & \vdots  \\
 \vdots & \ddots &  \ddots & \ddots  & 1 & 0  \\
0 &   &  \ddots &  0 & 0 & 1  \\
1 & 0 & ... & 0 & 0 & 0  \\
\end{array}
\right) \cr\cr\cr && g_2 = e^{i \pi (N-1)} \left(
\begin{array}{ccccccc}
e^{\frac{i 2\pi}{N}} &  & &  &   \\
 & e^{\frac{i 4\pi}{N}} & &  &    \\
 & & \ddots & & \\
& & & e^{\frac{i 2\pi(N-1)}{N}}  &   \\
& & & & 1  \\
\end{array}
\right), \label{Eq:LSM_Zn_Proj_Gen} \eeqn where $g_1$ only has non-zero
entries on a subdiagonal and the bottom left corner, and $g_2$ is
a diagonal matrix. It is straightforward to check that
\begin{align}
g_1^N = g_2^N = 1_{N \times N}, \ \ \ g_1 g_2 = e^{-i 2\pi /N} g_2
g_1.
\end{align}
We denote the elements of $\PSU(N)$ corresponding to $g_1$ and
$g_2$ as $\tilde{g}_1$ and $\tilde{g}_2$. Obviously,
$\tilde{g}_{1,2}$ are elements of order $N$. Since the phase
factor $e^{-i 2\pi /N} $ in the commutation relation between $g_1$
and $g_2$ is one of the center elements in $\SU(N)$,
$\tilde{g}_{1}$ and $\tilde{g}_{2}$ should commute in $\PSU(N)$.
Therefore, $\tilde{g}_1$ and $\tilde{g}_2$ generate a $Z_N \times
Z_N$ subgroup of $\PSU(N)$. We will focus on this subgroup in the
following. Notice that, a physical FR spin, which transforms
according to $g_{1,2}$ under this $Z_N \times Z_N$ subgroup of
$\PSU(N)$, can be viewed as a projective representation of $Z_N
\times Z_N$. In the classification of the projective
representation $H^2(Z_N \times Z_N, \U(1)) = Z_N$, the FR spins
actually correspond to the generating element in $H^2(Z_N \times
Z_N, \U(1))$. The AFR spins then correspond to the conjugate of
the FR spins in terms of projective representations of $Z_N \times
Z_N$.

When we restrict to the global internal symmetry $Z_N \times Z_N$
(which is a subgroup of $\PSU(N)$), we can apply the lattice
homotopy classification introduced in Ref.~\onlinecite{Po2017}. It
was proven for $1d$ and $2d$, partially proven for $3d$, and
conjectured for general dimensions that the generalized
Lieb-Schultz-Mattis (LSM) theorems will forbid the existence of
any featureless states on lattices of ``non-trivial lattice
homotopy class". In fact, the lattice homotopy classification
proposed in Ref.~\onlinecite{Po2017} also covers the cases with
continuous internal symmetry group. However, the proof of the
relations between non-trivial lattice homotopy classes and the
existence of generalized LSM theorems is less comprehensive for
the most general continuous symmetry group than for the general
Abelian finite group. Therefore, we will focus on the lattice
homotopy classification with Abelian finite group in this section.

For a lattice with $n$ FR spins on each site of the sublattice A
and $n$ AFR spins on each site of the sublattice B, we will refer
to it as the $(n,n)$-lattice. The fundamental-anti-fundamental
lattices can then also be referred to as the $(1,1)$-lattice. In
addition to the global internal symmetry, the lattice homotopy
classification depends on the choice of space group symmetry.
Let's specify the minimal space group symmetry for the
$(1,1)$-honeycomb, $(1,1)$-square and $(1,1)$-cubic lattices we
want to consider. For the $(1,1)$-honeycomb lattice, we want to at
least include the $C_3$ spatial rotation symmetry into
consideration. Therefore, the minimal choice of space group is the
wallpaper group $p3$ (No. 13). For the $(1,1)$-square lattice, we
want to at least consider the $C_4$ spatial rotation symmetry.
Therefore, the minimal choice of space group is the wallpaper
group $p4$ (No. 10). For the $(1, 1)$-cubic lattice, we want to at
least consider the symmetry of the point group $O$. Therefore, the
minimal choice of the 3D space group is $F432$ (No. 209). The
wallpaper group and 3D space group numbers can be found in
Ref.~\onlinecite{ITC}.

With the global $Z_N \times Z_N$ internal symmetry and the minimal
space groups symmetry given above, the $(1,1)$-honeycomb lattice
belongs to a non-trivial lattice homotopy class when $N$ is a
multiple of 3. Similarly, $(1,1)$-square and $(1,1)$-cubic
lattices are also non-trivial when $N$ is even. Therefore,
according to Ref.~\onlinecite{Po2017}, there are generalized LSM
theorems obstructing any featureless state compatible with the
global and space group symmetries on these lattices. Of course,
when we enlarge the $Z_N \times Z_N$ symmetry back to $\PSU(N)$,
such obstructions still exist.

Hence the analysis of lattice homotopy class also indicates that
{\it there is no featureless state with $\PSU(N)$ global symmetry
on the $(1,1)$-honeycomb lattice with $N$ being a multiple of 3,
or on $(1,1)$-square or cubic lattices with even integer $N$.}
These conclusions are completely consistent with those obtained
from the analysis in the previous sections.

One can perform a similar lattice homotopy analysis for SO$(2N)$
spin systems with spins carrying the vector representation with $N
\geq 1$. We focus on a $Z_2 \times Z_2$ subgroup of PSO$(2N)$.
When $N=4k$, we construct the SO$(4k)$ matrices \beqn g_1 = i
\sigma^y \otimes I_{2k \times 2k}, ~~~~ g_2 = \sigma^z \otimes
I_{2k \times 2k}, \eeqn and notice that \beqn g_1^2 = -1,~~ g_2^2
= 1,~~ g_1 g_2 = -g_2 g_1. \eeqn We denote the elements of
PSO$(4k)$ that correspond to $g_1$ and $g_2$ as $\tilde{g}_1$ and
$\tilde{g}_2$. Since $-I_{4k\times 4k}$ is a non-trivial center
element of $SO(4k)$, the elements $\tilde{g}_{1,2}$ generate a
$Z_2 \times Z_2$ subgroup of PSO$(4k)$. The vector representation,
which transforms according to $g_{1,2}$ under this $Z_2 \times
Z_2$ subgroup, can be viewed as a non-trivial projective
representation of $Z_2 \times Z_2$. If we restrict our attention
to this $Z_2 \times Z_2$ subgroup of PSO$(4k)$, we notice that a
square lattice with a SO$(4k)$ spin in the vector representation
per site and with the space group $p4$ belongs to a non-trivial
lattice homotopy class.

When $N=4k+2$, we construct the SO$(4k+2)$ matrices
\beqn
& g_1 =
\left(
\begin{array}{cccc}
\sigma^z & & & \\
& \sigma^z & & \\
& & i\sigma^y &  \\
& & & i\sigma^y \otimes I_{2(k-1) \times 2(k-1)}
\end{array}
\right),
\nonumber \\
& g_2 =
\left(
\begin{array}{cccc}
\sigma^x & & & \\
& i\sigma^y & & \\
& & \sigma^x &  \\
& & & \sigma^z \otimes I_{2(k-1) \times 2(k-1)}
\end{array}
\right) \eeqn which satisfy \beqn g_1^4 = g_2^4 =1,~~ g_1 g_2 =
-g_2 g_1. \eeqn By similar reasoning in the SO$(4k)$ case, we find
that the vector presentation of SO$(4k+2)$ can be viewed as a
non-trivial projective representation of a $Z_4 \times Z_4$
subgroup in PSO$(4k+2)$. In fact, the classification of projective
representation of $Z_4 \times Z_4$ is given by $H^2(Z_4 \times Z_4
, U(1)) = Z_4$ in which the vector representation belongs to the
``second" non-trivial class. When we consider the space group $p4$
and the $Z_4 \times Z_4$ subgroup of PSO$(4k+2)$ given above, we
notice that the square lattice with a spin in the vector
representation on each site also belongs to a non-trivial lattice
homotopy class, just like that case of SO$(4k)$.

Hence, we can conclude that {\it A SO($2N$) spin system with
vector representation on every site does not permit a featureless
gapped state on the square lattice. } This result completely
agrees with the analysis in the previous sections.

Lastly, we consider SO$(2N+1)$ spin systems with spinor
representations. SO$(2N+1)$ is the group of rotations in
$\mathbb{R}^{2N+1}$. Let $x_{1,2,...,2N+1}$ denote the $2N+1$ axes
of $\mathbb{R}^{2N+1}$. We'd like to focus on a $Z_2\times Z_2$
subgroup of SO$(2N+1)$ generated by the $\pi$-rotation in the
$x_1$-$x_2$ plane and the $\pi$-rotation in the $x_1$-$x_3$ plane.
The spinor representation of SO$(2N+1)$ can be viewed as a
non-trivial projective representation of this $Z_2 \times Z_2$
subgroup. When we consider the space group $p4$ and the $Z_2
\times Z_2$ subgroup of SO$(2N+1)$ given above, we notice that the
square lattice with a spin in the spinor representation on each
site belongs to a non-trivial lattice homotopy class. Therefore,
{\it a SO($2N+1$) spin system with spinor representation on every
site does not permit a featureless gapped state on the square
lattice. } Again, this statement is consistent with the analysis
given in the previous sections.

\section{Summary}

In this work we made connection between two seemingly different
subjects: the (generalized) Lieb-Shultz-Matthis theorem for a
$d-$dimensional quantum spin systems, and the boundary of
$(d+1)-$dimensional symmetry protected topological states with
on-site symmetries. This connection has led to fruitful results:
we identified a series of quantum spin systems that permit a
featureless spin state, as well as spin systems with a generalized
LSM theorem $i.e.$ spin systems that do not permit a featureless
spin state. The former cases correspond to trivial bulk states,
while the latter correspond to nontrivial SPT states in one higher
spatial dimensions. We have also tested and verified our
conclusions by other methods. For example we explicitly
constructed featureless tensor product spin states of those
systems whose corresponding $(d+1)-$dimensional bulk are trivial
states (most of this construction will be presented in an upcoming
paper~\cite{jianfuture}). We expect the main logic and method used
in this paper can be generalized to other related problems. For
example, one can study SU($N$) spin systems with more general
representations.

Zhen Bi and Cenke Xu are supported by the David and Lucile Packard
Foundation and NSF Grant No. DMR-1151208. Chao-Ming Jian's
research at KITP is supported a fellowship from the Gordon and
Betty Moore Foundation (Grant 4304). We thank Yi-Zhuang You for
helpful discussions. We also acknowledge a few upcoming
independent works which may overlap with part of our
results~\cite{maxfuture,ashvinfuture}.

\bibliography{sptlsm}

\end{document}